\documentclass[aps,prb,twocolumn,groupedaddress,floatfix,showpacs,showkeys,amsmath,amssymb]{revtex4-1}
\usepackage{graphicx}
\usepackage{bm}
\usepackage{hyperref}
\usepackage{amssymb}
\usepackage{amsmath}
\usepackage{bbm}
\usepackage{epsfig}
\usepackage{epstopdf}
\usepackage[usenames]{color}
\begin{document}
\title{Charge transfer and metallicity in LaNiO$_3$/LaMnO$_3$ superlattices}

\author{Alejandro Lopez-Bezanilla$^{1}$}
\email[]{alejandrolb@gmail.com}
\author{Louis-Fran\c cois Arsenault$^{2}$}
  \author{Anand Bhattacharya$^{1}$}
\author{Peter B. Littlewood$^{1}$ $^{3}$}
\author{Andrew J. Millis$^{2}$}

\affiliation{$^{1}$Materials Science Division, Argonne National Laboratory, Argonne, Illinois 60439, USA}
\affiliation{$^{2}$Department of Physics, Columbia University, New York, New York 10027, USA}
\affiliation{$^{3}$James Franck Institute, University of Chicago, Chicago, Illinois 60637, United States}

\begin{abstract}
Motivated by recent experiments, we use the $+U$ extension of the generalized gradient approximation to density functional theory to study superlattices composed of alternating layers of LaNiO$_3$ and LaMnO$_3$. For comparison we also study a rocksalt ((111) double perovskite) structure and bulk LaNiO$_3$ and LaMnO$_3$. A Wannier function analysis indicates that band parameters are transferable from bulk to superlattice situations with the exception of the transition metal d-level energy, which has a contribution from the change in d-shell occupancy. The charge transfer from Mn to Ni is found to be moderate in the superlattice, indicating  metallic behavior, in contrast to  the insulating behavior found in recent experiments, while the rocksalt structure is found to be insulating with a large Mn-Ni charge transfer.  We suggest a high density of cation antisite defects may account for the insulating behavior experimentally observed in short-period superlattices. 
\end{abstract}
\maketitle
 
 \section{Introduction}
The sensitive dependence of correlated electron properties on electron concentration and crystal structure has motivated the exploration of new systems that provide access to different regimes of the structure/concentration phase space. Oxide superlattices involving components with correlated electron properties are of particular interest because they present the possibility of controlled synthesis of correlated materials with specifically designed properties.\cite{Chakhalian14} Efficient exploitation of the new materials fabrication capabilities to establish a broadly effective ``materials by design'' capability will be enhanced by the development and validation of theoretical methods and physical/chemical understanding of the relation between structure and correlated electron properties. Improved understanding of ``transferability''--the extent, to which parameters  established by study of simpler system may be carried over to the description of a second, more complicated, situation--is important to this process, because to the extent that parameters are transferrable, intuition obtained from studies of model system and simple bulk compounds  can be used to guide synthesis and {\em ab initio} studies of complex structures. 

In this paper we address the validation and transferability issues in superlattices in the context of combinations of perovskite manganite (LaMnO$_3$ or LMO) and nickelate (LaNiO$_3$ or LNO) materials.  Individually these transition-metal oxide based materials exhibit remarkable physical properties related to the interplay of  magnetic, charge, orbital, and structural degrees of freedom, including colossal magnetoresistance, metal-insulator transitions and  orbital and spin-ordered states\cite{Imada98}. Chemical doping, pressure and  magnetic and electric fields have been employed to tune manganites and nickelates from one phase to another \cite{Tokura06,Kimura00,Middey16}.  Piamonteze and co--workers \cite{Piamonteze15} and Hoffman et al \cite{Hoffman16} have demonstrated that interesting magnetic structures can be induced  at interfaces between  LNO and LMO and also  also between La$_{2/3}$Sr$_{1/3}$MnO$_3$.

The present work is specifically motivated by results of Hoffman et al \cite{PhysRevB.88.144411}, who fabricated  superlattices of the chemical formula (LaNiO$_3$)$_n$(LaMnO$_3$)$_n$ and showed transport and optical evidence of a metal-insulator transition occurring as $n$ was decreased from 3 to 2. These authors further  used X-ray spectroscopy to show a thickness-dependent  transfer of electrons from Mn to Ni. Specifically, for a (LNO)$_2$/(LMO)$_2$ superlattice the cations exhibit spectra consistent with Mn$^{4+}$ and Ni$^{2+}$ oxidation states, very different from the  nominal Mn$^{3+}$ and Ni$^{3+}$ oxidation states observed in the corresponding bulk materials, and associated the electron transfer with the insulating behavior. 
 
In this paper we report results of density functional theory and density functional theory plus U (DFT+U) calculations performed to help understand the behavior of these materials. While DFT+U is a mean field method that does not capture the full complexity of correlated electron materials, it does provide reasonable estimates of basic physics such as charge transfer, allows for structural relaxation, and permits a detailed analysis of parameter transferability. We consider superlattices similar to those studied experimentally and for comparison also present results for  the cubic ABO$_3$ perovskite and the rocksalt A$_2$BB$^\prime$O$_6$  ((111) double pervoskite. We deal with idealized situations to get a sense of parameters and transferability and comment on the consequences of including more realistic details of the crystal structure.   Maximally localized Wannier functions (MLWFs) \cite{PhysRevB.56.12847} are employed to fit the band structures of the bulk compounds to  tight-binding models which parametrize the Hamiltonian description of each compound. Comparison of parameters obtained from the Wannier fits indicates that the on-site energy and intersite hopping parameters are transferable (meaning that they take the same value as in the bulk parent compounds), with one exception: the transition metal electronegativity is not transferable, but instead depends on the degree of charge transfer. These results indicate that insights from bulk materials can to a very large degree be carried over to the superlattice situation (making appropriate allowance for superlattice-induced changes in structure), but underscore the importance of an improved understanding of charge transfer and of the on-site energetics of the transition metal ions. 

Our results have implications for the interpretation of the experiments of Hoffman et al \cite{PhysRevB.88.144411}.  In the layered situation  the calculated  charge transfer is much less than the one electron per Mn found experimentally, even after accounting for possible lattice relaxation;  however  the charge transfer found for the rocksalt structure is  closer to the experimental value, suggesting that the structures fabricated by Hoffman et al have a high concentration of transition metal antisite defects, as predicted by previous work,\cite{Chen16}  so that the actual experimental situation may correspond more closely to the rocksalt structure. 

The rest of this paper is organized as follows. Section ~\ref{sec:model} presents the systems to be investigated and outlines the basic physics, section ~\ref{methods} presents the model and calculational methods and the energy bands that are the basic result. Section ~\ref{analysis} presents an interpretation of the results. Section ~\ref{sec:conclusions} is a summary and conclusion.

\section{Model\label{sec:model}}

\begin{figure}[htp]
\includegraphics[width=0.45\textwidth]{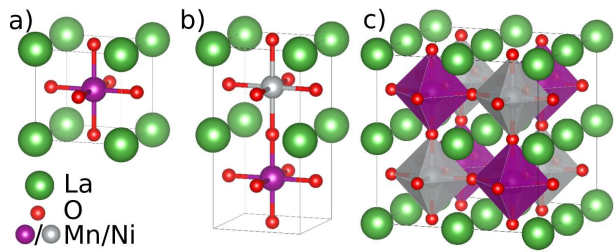}
\caption{ Model representation of the structures studied in this paper. a) cubic LaNiO$_3$ or LaMnO$_3$, b) multilayer of [LaMnO$_3$]$_1$/[LaNiO$_3$]$_1$, and c) rocksalt structure of LaMnO$_3$ and LaNiO$_3$. In panel (c)  where the octahedra represents the six-fold coordination of the metallic cations. }
\label{fig:structures}
\end{figure}

The systems studied are shown in Fig.~\ref{fig:structures}.  We consider the ideal cubic perovskite version of the bulk ``parent compounds" LaNiO$_3$ and LaMnO$_3$ (panels (a) and (b), a slightly idealized version of the superlattices studied experimentally (panel (c) of Fig.~\ref{fig:structures}), and  a rocksalt (111 double perovskite) structure with interpenetrating Mn and Ni sublattices (panel (d) of Fig.~\ref{fig:structures}). We idealize the structures to minimize the number of parameters describing the electronic physics, thus enabling a straightforward interpretation of the computations. 

We begin by describing the bulk ''parent compounds'' LaNiO$_3$ and LaMnO$_3$. LaNiO$_3$ (Figure \ref{fig:structures}a)  is the only one of the rare earth nickelate perovskites which remains metallic and non-magnetic down to the lowest temperatures (substituting another rare earth for La produces materials that have a low temperature insulating phase characterized by a two-sublattice breathing distortion and by antiferromagnetism).  LaNiO$_3$ forms in a slightly distorted version of the ideal ABO$_3$ perovskite structure, characterized by R$\bar{3}$c symmetry and a pseudocubic lattice parameter of 3.83 \AA. Here we approximate the structure as the simple cubic ABO$_3$ perovskite. Because we wish to compare to films grown on SrTiO$_3$ substrates we will choose the lattice constant to be equal to the SrTiO$_3$ pseudocubic lattice constant 3.95 \AA. The relevant electronic states in LaNiO$_3$ are antibonding combinations of Ni-d and O $2p\sigma$ orbitals. From the formal valence point of view, the Ni configuration is $d^7$  (t$^6_{2g}$-e$^1_{g}$) making the material  a representative spin-1/2,  orbitally degenerate  material.  However the highest-lying oxygen states are in fact believed to lie slightly higher in energy than the Ni $e_g$ levels, placing the material in the ``negative charge transfer'' class of materials \cite{Mizokawa00,Han11,Park12,Johnston14} so that the actual electronic configuration is closer to $d^8\bar{L}$, with Ni in the high-spin (t$^6_{2g}$-e$^2_{g}$) state and one hole on the oxygen (ligand) network. Theoretical calculations predict \cite{Chen13} and experiment confirms \cite{Cao16} that in LaNiO$_3$/LaTiO$_3$ superlattices, one electron is transferred from the LaTiO$_3$ layer to the LaNiO$_3$ layer,   changing the formal valence to a configuration similar to that of rocksalt NiO, a well known Mott/charge transfer insulator and producing insulating behavior. 

Bulk LaMnO$_3$ is an antiferromagnetic insulator; formal valence and Hund's coupling arguments indicate the $Mn$ configuration is high-spin $d^4$ with half-filled fully spin polarized $t_{2g}$-symmetry orbitals contributing an electrically inert $S=1$ ``core spin'' and the quarter filled $e_g$ manifold adding a potentially mobile $s=1/2$ carrier whose spin is strongly aligned to the core spin \cite{Anderson55,Millis95,Imada98} .  In bulk LaMnO$_3$,\cite{Kanamori60,Ellemans71} Jahn-Teller distortions involving alternating Mn-O bond lengths and  GdFeO$_3$-type checkerboard tilting of the oxygen octahedra lift the orbital degeneracy of the e$_g$ manifold, leading to an insulating ground state. Hole doping of LaMnO$_3$ reduces the tendency to Jahn-Teller order and the fully hole-doped end member SrMnO$_3$ is a cubic perovskite. Because we are interested situations in which charge transfer occurs, and because the Jahn-Teller ordering is in many cases  suppressed in manganite films \cite{Zhang15}, we study the cubic symmetry crystal structure (Figure \ref{fig:structures}a); again for comparison to films we use the bulk SrTiO$_3$ lattice constant of 3.95\AA.

\begin{figure*}[htbp]
 \centering
  \includegraphics[width=0.85\textwidth]{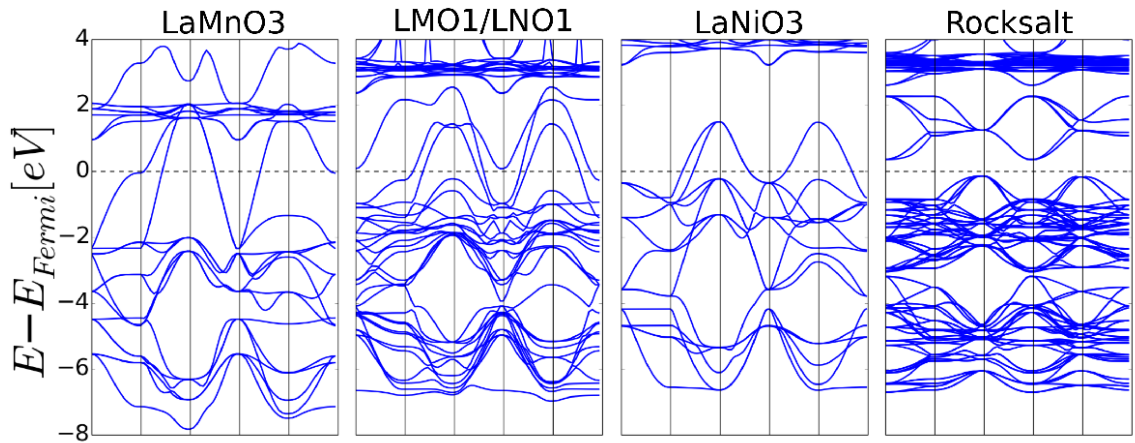}
 \caption{Majority spin bands of cubic LaMnO$_3$,  LaNiO$_3$,  the (001) (LaNiO$_3$)$_1$/(LaMnO$_3$)$_1$ superlattice and the rocksalt (111) La$_2$MnNiO$_6$ double perovskite. All calculations are performed for ferromagnetic ground states using the GGA+U method as described in the text.  
 The bands are plotted along the high-symmetry lines $\Gamma \rightarrow X \rightarrow R \rightarrow \Gamma \rightarrow M \rightarrow X$ of the Brillouin zone.}
 \label{fig2}
\end{figure*}
 \section{Method  \label{methods}}

A layered (LaNiO$_3$)$_n$/(LaMnO$_3$)$_n$ heterostructure may be formed by the periodic stacking of $n$ two-dimensional layers of both LaNiO$_3$ and LaMnO$_3$ in the ideal perovskite $(001)$ direction. Figure \ref{fig:structures}b shows the $n=1$ case, which we focus on in this paper for simplicity of interpretation. We note that this structure should show  maximal charge transfer effects.  To account for the strain induced by the substrate on which the experimental heterostructures were grown, the in-plane lattice parameters are fixed to the calculated bulk SrTiO$_3$ lattice constant of 3.95 and the out of plane lattice parameter to 2$\times$ 3.95\AA\ and rotation of the transition metal-oxygen octahedra were not allowed. We considered three cases for internal coordinates: the apical O atoms  at the same distance between the Mn and the Ni atoms, and displaced 0.2 and 0.4 \AA\  towards the Mn.

Finally we  studied a $A_2BB^\prime O_6$ double perovskite (rocksalt) structure (Figure \ref{fig:structures}c) in which  the six nearest cations to each Mn are Ni  and conversely. The $Mn$-$Ni$ distance was again taken to be equal to the SrTiO$_3$ pseudocubic lattice constant of 3.95\AA.  This structure is chosen to provide extra insight into the dependence of charge transfer on structure and also serves as a computationally tractable proxy for a superlattice  with a very high density of cation antisite defects.

We use the density functional +U (DFT+U) method, in which density functional band calculations are supplemented with on-site interactions among the transition metal d-orbitals. While DFT+U is a mean field method that does not fully capture the complexities of correlated electron physics, it does adequately represent the basic energetics of materials, in particular capturing correctly the basic energetics associated with charge transfer. An important issue in this method is the choice of interaction parameters  U and J and the corresponding double counting correction. According to previously reported results \cite{0953-8984-24-23-235602}, an on-site intra-d interaction term  U$_{Mn}$=4 eV and J$_{Mn}$=1.0eV, provides an adequate description of LaMnO$_3$ , successfully separating the higher-lying local minority spin t$_{2g}$ states from the local majority e$_g$ bands above a band gap which is likewise increased. When used with the experimental low temperature {\it Pbnm} structure this U$_{Mn}$ produces results consistent with experiment. The on-site interaction term for Ni has been chosen to U$_{Ni}$=6.0 eV J$_{Ni}$=1.0 eV  comparable to  values used in the literature \cite{PhysRevB.92.245109}.  

The fully localized limit (FLL) double-counting correction was employed. This double counting correction in effect compensates the Hartree shift of the transition metal d orbital, fixing the relative electronegativities of the transition metal and oxygen orbitals at about the values found in the density functional calculations.

The spin-polarized DFT calculations were performed using the projector augmented-wave method \cite{PhysRevB.50.17953} and the PBE-GGA exchange-correlation functional \cite{PhysRevLett.77.3865} with +U corrections in the Dudarev et al. scheme \cite{PhysRevB.57.1505} as implemented in the VASP code \cite{PhysRevB.48.13115,PhysRevB.54.11169,PhysRevB.59.1758}.  The rotationally invariant method with an effective U$_{eff}= U - J$ was employed. The electronic wave-functions were described using a plane-wave basis set with an energy cutoff of 500 eV. Atomic positions were fixed in the cubic or supercell  geometries described in the previous section, with a lattice parameter of 3.95 \AA. A 10$\times$10$\times$10 k-point sampling in $\Gamma$-centered cubic cells was used. The number of k-points was decreased proportionally as the number of cells increased in the supercell. 

The calculations presented here  consider ferromagnetic ground states, because the small unit cell size enables a clear interpretation of the band structure. To analyse the band structures we fit the bands using  maximally localized Wannier function \cite{PhysRevB.56.12847} methods as implemented in the VASP  Wannier90 code\cite{Mostofi2008685}. We  chose an energy window encompassing the p-d band complex and projected the Bloch functions onto p-symmetry orbitals centered on the O sites and d-symmetry orbitals centered on the transition metal sites then minimized the MLWF spread. The resulting Wannier bands were in excellent agreement with the VASP bands. From the Wannier fits we then obtained on-site energies for the  transition metal d and oxygen p levels, as well as d-p hopping amplitudes. 

We determined the charge transfer by comparing the integral of the valence band charge density (from the VASP output file CHGCAR) over the volume of each cubic sub-cell of the heterostructure to the corresponding values obtained from obtained on the individual LMO and LNO cells in the cubic and rocksalt structures.

\section{Results: Energy Bands and Density of States}

The four panels of Figure \ref{fig2} show the majority-spin energy bands of  cubic LaMnO$_3$, cubic LaNiO$_3$,the 1/1 (001) superlattice and the rocksalt (double perovskite) La$_2$MnNiO$_6$.

Cubic LaMnO$_3$ is predicted to be metallic (obtaining insulating behavior requires including both the Jahn-Teller distortion and octahedral rotations). The bands that cross the fermi level are about 4 eV wide and are of primarily Mn $e_g$ origin (hybridized with O$_{p\sigma}$ orbitals lying about 2eV below the Fermi level). These bands contain one electron shared between the two orbitals. In LaNiO$_3$ a similar situation occurs, but the rare earth perovskite nickelates are ``negative charge transfer materials'' in which the  O$_{2p\sigma}$ orbitals lie in fact at a slightly higher energy than the Ni $e_g$ orbitals implying,  as will be seen in more detail below, that  the bands crossing the Fermi level are p-d $e_g$-symmetry hybrids with majority p character. The calculated band structure for the superlattice reveals a metallic state with two relatively wide bands that cross the Fermi level. As will be seen in more detail below, one of these bands arises from the Mn $d_{x^2-y^2}$ e$_g$ state and the other from the Ni p-d hybrid state of the same symmetry. Also visible just above the fermi level is an unoccupied band mainly derived from the Mn-Ni antibonding combination of $d_{3z^2-r^2}$ e$_g$ symmetry states (the bonding combination lies several eV below the Fermi level and is obscured by the many other orbitals in this energy range).  

The metallic behavior of the superlattice may be understood as arising because the in-plane bands are only weakly affected by the superlattice formation. The wide energy range over which the bands disperse inhibits complete charge transfer. In fact, the net charge transferred from the Mn plane to the Ni plane is 0.4 e/unit cell.  The calculated bands for the rocksalt structure reveal an insulator, essentially because in this structure both $e_g$ orbitals on the $Mn$ site mix strongly with the $e_g$ orbitals on the $Ni$ site, allowing splitting of all bands. The narrower bandwidth also promotes a larger charge transfer, approximately 0.7 e/unit cell. 

\begin{figure}[htp]
 \centering
  \includegraphics[width=0.45 \textwidth]{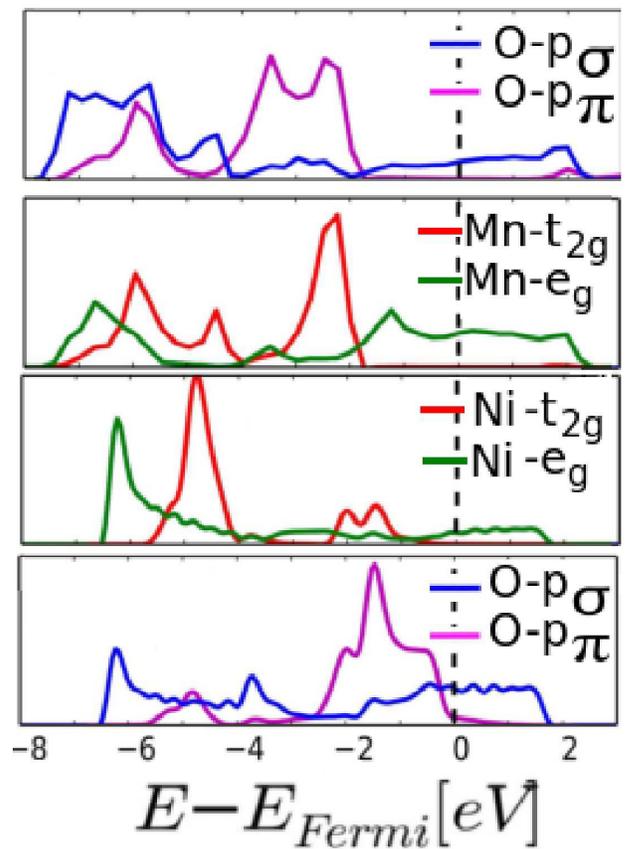}
 \caption{Projection of the majority spin density of states of ferromagnetic cubic LaMnO$_3$ and LaNiO$_3$ onto the transition metal (Mn or Ni) d orbitals and onto oxygen p orbitals. The  $e_g$-symmetry and $t_{2g}$-symmetry  orbitals are plotted separately (color online) as are the oxygen $p_\sigma$ (hybridizing with the transition metal $e_g$) and $p_\pi$ (not hybridizing with the transition metal $e_g$ orbitals).}
  \label{fig:cubicpdos}
\end{figure}

\begin{figure}[htp]
 \centering
  \includegraphics[width=0.45 \textwidth]{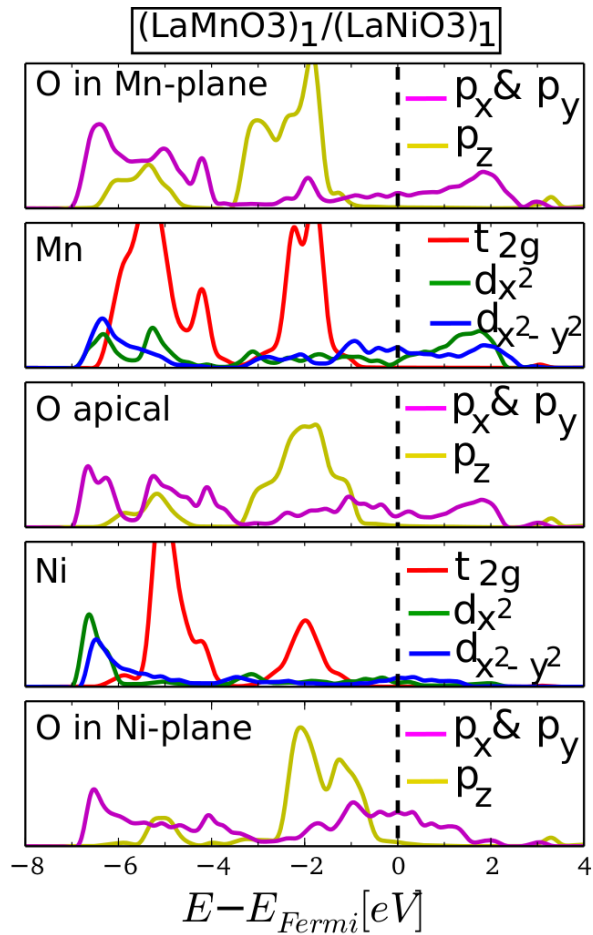}
 \caption{Projection of the majority spin density of states of the  ferromagnetic (LaNiO$_3$)$_1$/(LaMnO$_3$)$_1$ superlattice onto the transition metal (Mn or Ni) d orbitals and onto oxygen p orbitals.  The  two $e_g$ orbitals are plotted separately, but all three $t_{2g}$ orbitals are summed together. The oxygen $p_\sigma$ (hybridizing with the transition metal $e_g$) and $p_\pi$ (not hybridizing with the transition metal $e_g$ orbitals) distinguished.}
  \label{fig:slpdos}
\end{figure}

\begin{figure}[htp]
 \centering
  \includegraphics[width=0.45 \textwidth]{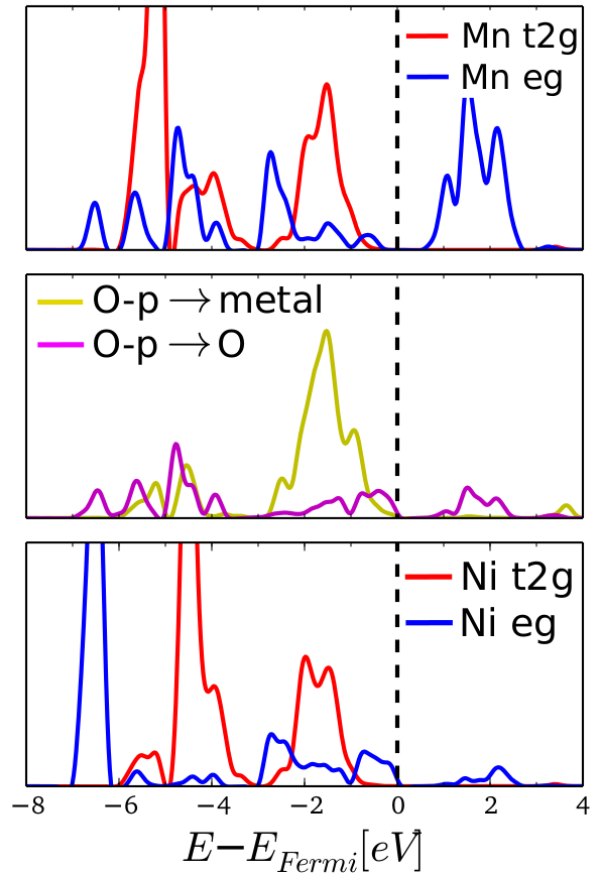}
 \caption{Projection of the majority spin density of states of ferromagnetic rocksalt A$_2$BB$^\prime$O$_6$ onto $e_g$ and $t_{2g}$ orbitals are plotted separately. The oxygen $p_\sigma$ (hybridizing with the transition metal $e_g$) and $p_\pi$ (not hybridizing with the transition metal $e_g$ orbitals) distinguished. }
  \label{fig:rocksaltpdos}
\end{figure}

To elucidate the physics revealed by our calculated band structures we  consider the  electronic density of states, projected onto particular symmetry states of particular atoms. Fig.~\ref{fig:cubicpdos} shows the orbitally projected densities of states for the cubic parent materials.  From the top two  rows we see that in LaMnO$_3$ the near Fermi surface states are  $e_g$ symmetry antibonding d-p hybrids, with majority d character.. The corresponding bonding states are visible as the peak centered at $\approx -7eV$. Comparison of the relative amplitude of bonding and antibonding states in the d and p partial DOS confirms that the $e_g$ d-states lie above the p-states in this material. The $t_{2g}$-symmetry states are fully filled, and couple  less strongly to the oxygen orbitals, as revealed by their smaller peak width and smaller bonding-antibonding splitting. 

Turning now to the lower two panels we see that for LaNiO$_3$  the Ni p-d manifold is somewhat narrower in energy than the p-d manifold in the Mn compound, reflecting a weaker p-d hybridization in the Ni material. The relative weights of the  bonding and antibonding peaks and the relative magnitudes of the p and d components indicate that in LaNiO$_3$ the d-states lie lower than the oxygen states, as expected from the greater electronegativity of Ni relative to Mn.  As in the Mn compound the Ni $t_{2g}$ states are fully filled and rather less strongly hybridized to the oxygen than the $e_g$ states. 

We now turn to the density of states of the superlattice, shown in Fig.~\ref{fig:slpdos}.  The top panel shows the density of states projected onto the oxygen states lying in the $Mn-O$ plane, and the second panel shows the DOS projected onto the Mn states. Comparison to the corresponding panels of Fig.~\ref{fig:cubicpdos} shows that the superlattice-induced changes to the Mn-plane O orbitals and the Mn $d_{x^2-y^2}$ orbital are negligible. A splitting of the Mn $d_{3z^2-r^2}$ orbital is evident (leading e.g. to the suppressed DOS at the Fermi level) with corresponding shifts to the ``apical'' (in-between Mn and Ni planes) oxygen orbital.  Similarly, examination of the projection of the density of states onto the Ni and Ni-plane O orbitals reveals negligible changes to the planar orbitals; the small weight of the Ni d orbitals near the fermi level makes changes in the Ni $d_{3z^2-r^2}$ orbitals difficult to discern

These results confirm that  one of the bands that crosses the Fermi level is primarily of Mn d$_{x^2-y^2}$ character and the other primarily of O character but within the $Ni-O$ plane,  with the antibonding portion of the  Mn $d_{3z^2-r^2}$ orbital and the $O_{p_z}$ orbital split, with one portion pushed up in energy above the Fermi level by backscattering associated with the breaking of translational symmetry in the superlattice.    

Finally, in Fig. ~\ref{fig:rocksaltpdos} we show the partial densities of states for the rocksalt structure. The $Mn$ d orbitals are now to a large degree pushed up above the Fermi level and the Ni d orbitals and oxygen orbitals are submerged below, leading to insulating behavior and a larger degree of charge transfer. 

\section{Parameter Transferability: Wannier Analysis \label{analysis}}

The panels of Fig. ~\ref{fig2} show that as the unit cell  becomes larger, the number of  energy bands increases, as does the band overlap, so the physics becomes more difficult to analyse. The bands would become even more complicated if octahedral rotations or antiferromagnetism were considered. For this reason, an important issue in analysing the physics of nontrivial situations is parameter  `transferability': specifically, whether one can use parameters obtained in simpler situations to model more complicated ones. We have examined the transferability issue in the systems we consider by performing a Maximally Localized Wannier Function (MLWF) analysis of our calculated bands for each of the structures. The MLWF analysis can be thought of as an unbiassed determination of tight binding parameters; comparison of the parameters obtained for different situations then provides an estimate of transferability. 

We find that the hopping amplitudes (inter-orbital matrix elements of the DFT+U Hamiltonian) are the same within a few percent for all structures (e.g. the Mn $e_g$-O$2p_\sigma$ hopping is 1.7 and the Ni $e_g$-O$2p_\sigma$ hopping is 1.3eV). We conclude that the hopping parameters are transferable.  

The on-site energies (orbital-diagonal matrix elements of the DFT+U Hamiltonian) exhibit variation among systems. One contribution to the variation is changes to the combination of  Madelung energies associated with different chemical environments as well as any electric fields arising from charge transfer (this is essentially the intrinsic contribution to the work function difference that gives rise to band-bending at interfaces). For example, the photoemission data reported by Yoshimatsu et al for  SrVO$_3$/SrTiO$_3$ quantum wells \cite{Yoshimatsu10,Yoshimatsu11}, indicates a $\sim 0.9eV$ shift in the oxygen energy from the one material to the other.  A second contribution to the variation is a change of physics, in particular a variation of the p-d energy level splitting, which plays a crucial role in the physics of transition metal oxides \cite{Zaanen85,Dang14}.

\begin{table}[htp]
\caption{Orbital energies obtained from Maximally Localized Wannier analysis for band structures computed with $U_{Mn}=4$, $U_{Ni}=6$ eV. $J=1eV$ and measured relative to the Fermi level.}
\begin{center}
\begin{tabular}{|c|c|clcl|}
\hline \hline
Orbital&LaMnO$_3$&$SL$&RS&LaNiO$_3$&\\
\hline

Mn$_{ z^2}$&-2.2&-1.7&-1.4&--&\\
Mn$_{ x^2-y^2}$&-2.2&-2.0&-1.4&--&\\
O:Mn-plane&-4.6&-3.8&-3.0&-&\\
O:Apical&-4.6&-3.4&-3.0&-2.7&\\
Ni$_{ z^2}$&--&-4.2&-4.0&-3.6&\\
Ni$_{ x^2-y^2}$&--&-3.9&-4.0&-3.6&\\
O:Ni-plane&--&-3.0&-3.1&-2.7&\\
\hline
\end{tabular}
\end{center}
\label{table1}
\end{table}

Table ~\ref{table1} presents the energies of the Mn and Ni $e_g$ symmetry orbitals as well as  the energies of the $p_{\sigma}$ orbitals on different oxygen sites, distinguishing (for the layered structure) between oxygen states in the Mn or Ni plane and the apical oxygen sites that bridge the Mn and Ni planes.  The `band-bending' or work function differences  are clearly visible as differences between the energy of the O in the MnO$_2$ and NiO$_2$ planes. In addition, a change in the relative energies of the oxygen p and transition metal d levels is  apparent.

For cubic LaMnO$_3$  we have $E^{Mn}_{e_g}-E^O_{2p\sigma}\approx$ 2.4eV while for LaNiO$_3$  we find $E^{Ni}_{e_g}-E^O_{2p\sigma}\approx$ -0.9 eV, reflecting the negative charge transfer nature of the Ni compound.  In the rocksalt structure we find $E_{e_g}^{Mn}-E^O_{2p\sigma}\approx$ 1.6 eV, about 0.6eV {\em less} than in  bulk LaMnO$_3$ while $E_{e_g}^{Ni}-E^O_{2p\sigma}\approx$ -1.0 eV, essentially unchanged from  bulk LaNiO$_3$. In the superlattice we see that the energy difference between the average of the energies of the two Mn orbitals and the energy of the in-plane oxygen $\sigma$ orbital $E_{e_g}^{Mn}-E^O_{2p\sigma}\approx$ 2.0 eV. We thus conclude that the charge transfer energy is {\em not} transferable. The  fact that the change is larger in the rocksalt than in the superlattice structure suggests that the d-valence makes an important contribution to the p-d level splitting. 

Interestingly the p-d splitting on Ni site is almost material independent. We attribute this to the negative charge transfer nature of the material, which  implies (as can be seen from the density of states plot) that the near Fermi surface states are of primarily oxygen character. Thus in the approximation used here the actual charge density on the Ni sites changes only slightly (loosely speaking the Ni-O plane goes from $d^8{\bar L}$ to $d^8$ as charges are added) and the mean valence of each of the 3 oxygens changes by less than e/3, so the oxygen Hartree shift is minimal, explaining why the $\varepsilon_d-\varepsilon_p$ changes only slightly.

\section{Comparison to experiment}

The essential features of the experiment of Hoffman et al \cite{PhysRevB.88.144411} are that for (LaMnO$_3$)$_n$/(LaNiO$_3$)$_n$ superlattices with $n\leq 2$ the ground state was insulating and an almost complete charge transfer of one electron occurred from the Mn to the Ni. These behaviors are not reproduced by the calculations, which predict that even for the $n=1$ superlattice we have metallic behavior with $\sim 0.4e$ charge transfer. The two effects are closely related: the planar orbitals ($d_{x^2-y^2}$ and in-plane $p_\sigma$) are only minimally affected by superlattice formation, and the resulting bandwidths are so large that the moderate Mn-Ni electronegativity difference cannot empty out the Mn d-bands. While our calculations are based on the DFT+U approximation, a Hartree approximation to a complicated many-body situation, we believe that correcting the deficiencies of this approximation are unlikely to change the basic theoretical prediction because the conclusion arises from eV-scale energetics of different valence states, which are well captured by DFT/Hartree approximations.  

The results presented here were derived from idealized cubic-type  structures, however we have also investigated the consequences of changing the structure.  Changing the lattice parameter of the cells to those of LNO or LMO only changes the dispersion of the electronic states. Modifying the spin orientation on the Mn and Ni sites to the anti-ferromagnetic coupling does not open a full band gap in any of the heterostructures  considered: the $d_{x^2-y^2}$ band remains metallic in all cases. Relaxing internal coordinates also does not change the physics fundamentally. Octahedral rotations do affect bandwidths, but because the in-plane lattice constant of the superlattice is fixed by the SrTiO$_3$ substrate to be 3.95\AA\, substantially {\em larger} than the lattice constant of La/SrMnO$_3$ (metallic at all relevant Sr concentrations), we expect that octahedral rotations will not substantially change our results. 

Another possibility is that the electron transfer and differing hybridization strengths will cause motion of the apical oxygen away from the Ni and towards the Mn. However relaxing the internal coordinates only leads to a $\sim0.1\AA$ shift in the apical O position, too small to change the energetics significantly.  We also investigated variant structures in which the apical O is displaced by 0.2\AA\ and $0.4\AA$ toward the Mn. The charge transfer increases to $0.53$ and $0.7$ electrons respectively, but the $d_{x^2-y^2}$-derived orbitals remain partially occupied and the superlattice remains metallic. 

Interestingly, the rocksalt structure has a similar $0.7e$ charge transfer, but is insulating, essentially because the backscattering associated with the Mn-Ni alternation opens a band gap.  We conclude that the insulating behavior requires disruption of the in-plane Mn-O and Ni-O networks and speculate that a high density of Mn/Ni atisite defects in the near interface layers could disrupt the in-plane networks as well as promoting larger charge transfer, thus  producing physics similar to that found in the calculations of the rocksalt structure. In this regard it is interesting to note that evidence of substantial Ni-Mn intermixing, especially for the near-substrate layers, has been very recently reported in a closely related superlattice.\cite{Kwon17}

\begin{figure}[htp]
 \centering
  \includegraphics[width=0.45 \textwidth]{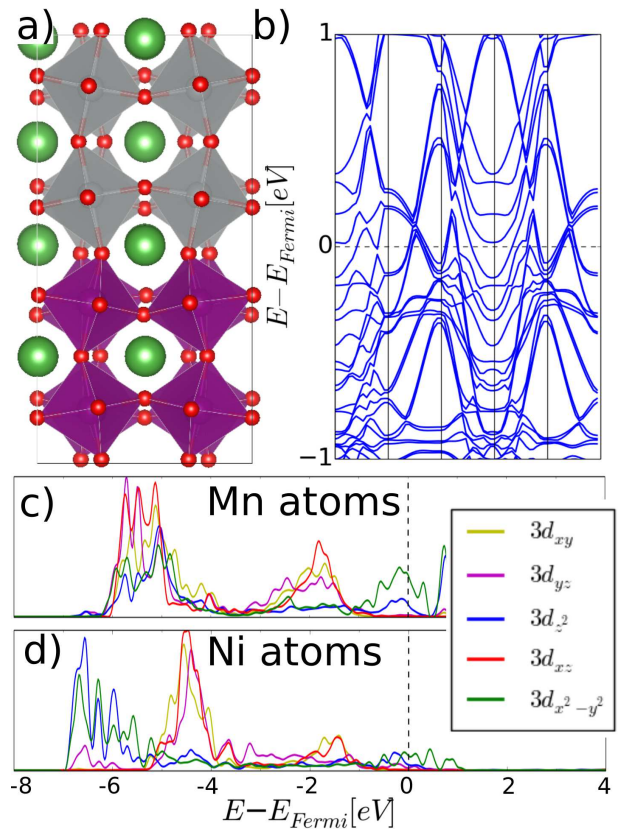}
 \caption{a) Schematic representation of the relaxed cubic (LaNiO$_3$)$_2$/(LaMnO$_3$)$_2$ heterostructure. b) Band structure of the majority spin. The projected DoS (PDoS) onto the d orbitals of the two types of nonequivalent Mn atoms is plotted in c) and d).}
 \label{fig5}
\end{figure}

Finally we consider the superlattice containing many of the possible degrees of freedom of the actual (LaNiO$_3$)$_2$/(LaMnO$_3$)$_2$ multilayer. A 2$\times$2$\times$4 undistorted structure of LMO is coupled to the same type of LNO structure. Generally, if a Mn cation is in 4+ oxidation states, one should expect no Jahn-Teller distortion to occur in the Mn sublattice. But when the internal coordinates are relaxed, allowing the system to find the optimal geometry, the LNO follows the distortions imposed by the LMO (see Fig. \ref{fig5}), again consistent with a small to moderate charge transfer.  The band diagram in Fig. \ref{fig5} shows the ground state corresponding to the Mn sublattice with an antiferromagnetic spin alignment for Mn atoms in parallel planes, and ferromagnetic spin alignment for the Ni atoms. It shows that the hybrid structure is metallic with numerous bands crossing the Fermi level. A projected DOS analysis demonstrates that the e$_g$ orbitals lead the metallic properties of the structure, with the in-plane d$_{x^2-y^2}$ orbitals contributing the most.

\section{Conclusions\label{sec:conclusions}} 
First-principles calculations of model  LaMnO$_3$ and LaNiO$_3$ structures reveal that the basic physics of the superlattice situation is controlled by geometry, which determines hopping amplitudes and thus band structures, and by relative electronegativity of the transition metal ions, which determines charge transfer. Our calculations strongly suggest that in the ideal superlattice,  the wide $d_{x^2-y^2}$ symmetry bands are essentially unaffected by superlattice formation and  remain metallic because the Mn-Ni electronegativity difference is smaller than the bandwidth, so the Mn-derived bands remain partly occupied and the Ni-derived bands remain partly empty. This should be contrasted to the LaITiO$_3$LaNiO$_3$ superlattices studied previously \cite{Chen13}, in which the Ti-Ni electronegativity difference is large enough  and the Ti $d_{xy}$ band is narrow enough that nearly complete charge transfer occurs. 

An analysis based on maximally localized Wannier functions demonstrates that transition metal/oxygen hybridization amplitudes and oxygen site energies are ``transferable'', so that parameters derived from calculations performed in simple geometries can be used in phenomenological models of more complex situations. However, the transition metal d-oxygen p level splitting is not transferable: it is found to depend on the charge transfer, essentially because the transition metal electronegativity depends on d-occupancy. 

For the LMO/LNO system that motivated our study, we found that within our theoretical framework the only viable explanation for the observed insulating behavior of thin superlattices was a large concentration of antisite defects. Some experimental evidence for this possibility has recently been reported \cite{Kwon17}. As a possible alternative, we note that it is conceivable that some other mechanism (e.g. thermal fluctuations  at higher temperatures)  might lead to octahedral rotations of much larger amplitude than considered here, perhaps reducing the bandwidth in the layered structures enough to eliminate the overlap between bands, thereby allowing a larger charge transfer leading to insulating behavior. Experimental investigation of antisite defects and of octahedral rotations would be valuable. 

We suggest directions for future research. First, the charge transfer depends on the difference in electronegativity. The DFT+U approach used here provides an approximation to the electronegativity difference. Further studies of these issues by other methods would be valuable. Second, here parameter transferability has been investigated in one particular material system. We conjecture that the main result (transfer of all parameters except for the p-d level splitting) will extend to all transition metal oxides heterostructures, but  further investigation would be desirable.  Perhaps most importantly, it appears from our results that  the difference between the energies of the p and d orbitals of the Mn cation is not transferable between structures. We conjectured that intra-d Hartree energy plays an important role in the non-transferability.  Further studies of this basic and  important effect would be desirable.

\section{Acknowledgments}   
This work was supported by the U.S. Department of Energy, Office of Science, Basic Energy Sciences, Materials Science and Engineering Division. We acknowledge the computing resources provided on Blues and Fusion, the high-performance computing clusters operated by the Laboratory Computing Resource Center at Argonne National Laboratory.

%

\section{Appendix: Wannier analysis}
\begin{figure}[htp]
 \centering
  \includegraphics[width=0.5 \textwidth]{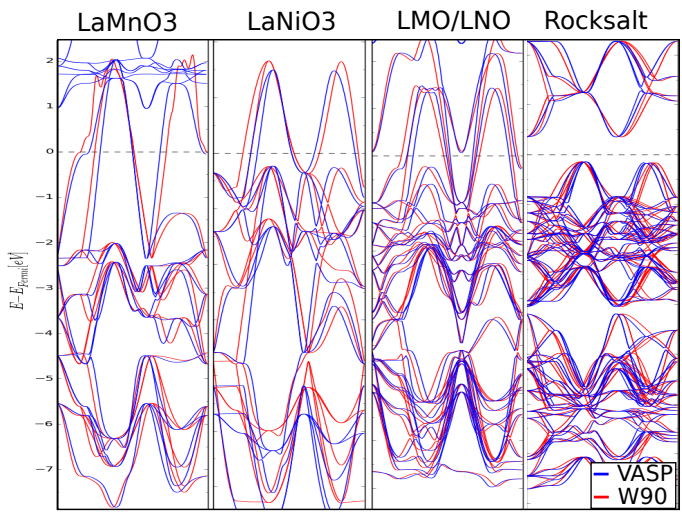}
 \caption{Comparison of the first-principles obtained band structure  with the Wannier90 interpolated band structure for cubic LaMnO$_3$, cubic LaNiO$_3$, (LaNiO$_3$)$_1$/(LaMnO$_3$)$_1$ heterostructure, and anti-ferromagnetic rocksalt.
 }
 \label{fig:wanniercomparison}
\end{figure}

To obtain model Hamiltonians we resort to the the maximally localized Wannier functions (MLWFs)\cite{PhysRevB.56.12847}, obtained using the Wannier90 code\cite{Mostofi2008685} from the first-principles ground state obtained with VASP.  By projecting onto p and d orbitals and minimizing the MLWF spread, the band structure obtained using the first-principles approach compared to the Wannier interpolation were in excellent agreement. Fig. \ref{fig:wanniercomparison} shows a comparison for the 4 structures of interest.

\end{document}